\documentstyle[12pt,rsfs]{article}

\title{Ewald summation technique for \\
       interaction site models of polar fluids}

\author{\sc Igor~P.~Omelyan \\ [1.5ex]
{\small \em Institute for Condensed Matter Physics,
            National Ukrainian Academy of Sciences} \\ [-8pt]
{\small \em 1~Svientsitsky St., UA-290011 Lviv, Ukraine \thanks
           {E-mail: nep@icmp.lviv.ua}} \\
\date{}}

\newcommand{\bms}[1]{\mbox{\boldmath $#1$}}
\newcommand{\bvs}[1]{\mbox{\scriptsize\boldmath $#1$}}
\newcommand{\scs}[1]{_{\stackrel{\ }{#1}}}

\begin{document}

\setlength{\abovedisplayskip}{18pt plus4pt minus6pt}
\setlength{\belowdisplayskip}{\abovedisplayskip}
\setlength{\abovedisplayshortskip}{12pt plus2pt minus4pt}
\setlength{\belowdisplayshortskip}{\abovedisplayshortskip}

\maketitle

\vspace{1cm}

\begin{abstract}

A computer adapted fluctuation formula for the calculation of the
wavevec\-tor- and frequency-dependent dielectric permittivity for
interaction site models of polar fluids within the Ewald summation
technique is proposed and applied to molecular dynamics simulations
of the TIP4P water. The formula is analyzed and optimal parameters
of the Ewald method are identified. A comparison of the obtained
results with those evaluated within the reaction field approach
is made.

\vspace{0.5cm}


\vspace{0.1cm}


\end{abstract}

\newpage

\section{Motivation}

\hspace{1em}  In order to achieve a macroscopic behaviour for investigated
quantities in computer experiment based on the observation of finite systems,
it is necessary to reduce the influence of surface effects to a minimum. This
is especially important for polar systems with the long-range nature of
interactions. Excluding the surface effects in simulations can be performed
within either the reaction field (RF) [1--5] or Ewald summation [6--10]
techniques. Now an equivalence between these techniques has been established
for models of point dipoles and proper calculations can be made within either
method [11, 12]. The explicit consideration of a finite-size medium lead to
computer adapted fluctuation formulas [11--17] which allow one to calculate
boundary free values for the dielectric constant on the basis of dipole
moment fluctuations obtained in simulations. These formulas differ
considerably with respect to those known from the theory of macroscopic
systems even if the Ewald method is used [11]. Details of the summation
must be taken into account explicitly in order to obtain correct values
for the bulk dielectric constant.

Previously [18--20], the standard RF of point dipoles (PDRF) [3] has been
applied to investigate more realistic, interaction site models (ISMs) [21]
of polar fluids. The PDRF, however, being exact for point dipole models,
may not be necessarily applicably to interpret simulation results for
arbitrary systems [5].
Recently, it has been shown by actual calculations
for a MCY water model that uncertainties for the dielectric quantities are
significant if the PDRF is used in computer simulations of ISMs and an
alternative scheme, the interaction site reaction field (ISRF) geometry,
has been proposed [22]. At the same time, there is not such an approach
concerning the entire wavevector and frequency dependence for the dielectric
permittivity of ISMs within the Ewald geometry. The main attention of
previous simulations [23--31] was directed to study the dielectric
properties in the static limit or at zero and small wavevector values.
Moreover, the macroscopic fluctuation formulas have been used in the
simulation results without taking into account details of the Ewald
summation.

In the present paper we apply the Ewald technique for treating Coulomb
interactions in ISMs. The paper is organized as follows. A fluctuation
formula suitable for the calculation of the wavevector- and
frequency-dependent dielectric constant is derived in Sec.~2 and optimal
values of the Ewald parameters are determined there. The results of
molecular dynamics simulations of the TIP4P water for time correlation
functions related to dielectric polarization are presented in Sec.~3.
These results are compared with those computed within the ISRF geometry.
Concluding remarks are given in Sec.~4.

\vspace{12pt}

\section{Ewald summation for ISMs}

\hspace{1em}  Consider a polar fluid with $N$ molecules composed of $M$
interaction sites which are confined in a volume $V$. The microscopic
electric field created by the molecules at point $\bms{r}$ and time $t$
can be presented as $\bms{\hat E}(\bms{r},t) = \displaystyle \int_V
\bms{D}(\bms{r}-\bms{r}') \hat{Q}(\bms{r}',t) {\rm d} \bms{r}'$, where
$\hat Q(\bms{r},t)=\sum_{i=1}^N \sum_{a=1}^M q \scs{a} \delta (\bms{r}-
\bms{r}_i^a(t))$ is the microscopic charge density, $\bms{r}_i^a(t)$ and
$q \scs{a}$ denote the position and charge, respectively, of site $a$
within the molecule $i$ and $\bms{D}(\bms{\rho})=-\bms{\nabla} \ 1/\rho$
is the operator of the Coulomb interactions.

Obviously, the field $\bms{\hat E}(\bms{r},t)$ for infinite systems $(N, V
\to \infty)$ can not be reproduced exactly in computer experiment which
deals with a finite, as a rule, cubic volume $V=L^3$, where $L$ is the
length of the simulation box edge. However, using the lattice summation, a
macroscopic behaviour can be achieved considering the interactions between
sites within the basic cell as well as an infinite lattice of its periodic
images (the periodic boundary convention). This can be interpreted as an
effective interaction which involves only the sites in the basic cell and
characterized by a modified operator $\bms{\cal D}(\bms{\rho})=\sum_{\bvs{n}}
\bms{D}(\bms{\rho}+\bms{n} L)$, where the summation is extended over all
vectors $\bms{n}$ with integer components. It is more convenient to represent
the lattice sum in a form, proposed by Ewald and Kornfeld (EK) [6], namely,
$\bms{\cal D}(\bms{\rho})=\bms{D}_1(\bms{\rho})+\bms{D}_2(\bms{\rho})$, where
\begin{equation}
\bms{D}_1(\bms{\rho})=\sum_{0 \le |\bvs{n}| \le {\cal N}}
\bms{D}(\bms{\rho}+\bms{n} L) \Big\{ {\rm erfc}(\eta |\bms{\rho}+\bms{n} L|+
\frac{2 \eta}{\sqrt{\pi}} |\bms{\rho}+\bms{n} L|
\exp(-\eta^2 |\bms{\rho}+\bms{n} L|^2) \Big\}
\end{equation}
is a sum in real coordinate space, while
\begin{equation}
\bms{D}_2(\bms{\rho})=\frac{1}{V} \sum_{0 < |\bvs{k}| \le k_{\rm max}}
\bms{D}(\bms{k}) \exp(-k^2/4\eta^2+{\rm i} \bms{k\!\cdot\!\rho})
\end{equation}
corresponds to summation over wavevectors $\bms{k}=2 \pi \bms{n}/L$ of the
reciprocal lattice space and $\bms{D}(\bms{k})=\displaystyle \int {\rm d}
\bms{r} \, {\mbox{\large e}}^{-{\rm i} \bvs{k\!\cdot\!\rho}} \bms{D}(\bms
{\rho}) = -4\pi {\rm i} \bms{k}/k^2$ is the spatial Fourier transform of
$\bms{D}(\bms{\rho})$. For the idealized summations $({\cal N} \to \infty,
k_{\rm max} \to \infty$), the total sum of (1) and (2) is independent on the
parameter $\eta$. The main advantage of the EK representation lies in the
fact that values for $\eta$ can be found in such a way that the both sums,
$\bms{D}_1$ and $\bms{D}_2$, converge very quickly and may be truncated after
a finite number of terms. If the parameter $\eta$ is chosen sufficiently
large, we can restrict ourselves to a single term $({\cal N}=0)$ in the
real space sum, corresponding to the basic cell to which toroidal boundary
conditions are applied, and, additionally, to the spherical truncation
$|\bms{\rho}| \le R$, where $R \le L/2$.

In such a case, taking the Fourier transforms of (1) and (2), after some
algebra one obtains $\bms{D}_1(\bms{k})=-4\pi{\rm i} D_1(k) \bms{k}/k^2$
and $\bms{D}_2(\bms{k}) = -4\pi{\rm i} D_2(k) \bms{k}/k^2$, where
\begin{equation}
D_1(k)=\int \limits_0^R k j\scs{1}(k\rho) \Big( {\rm erfc}(\eta \rho)+
\frac{2 \eta}{\sqrt{\pi}} \rho \exp(-\eta^2 \rho^2) \Big) {\rm d} \rho \ ,
\end{equation}
$D_2(k)=\exp(-k^2/4\eta^2)$ if $0 < k \le k_{\rm max}$ and $D_2(k)=0$
otherwise and $j\scs{1}(z)= \sin(z)/z^2 - \cos(z)/z$ denotes the spherical
Bessel function of first order. Then the Fourier transform of the electric
field is
\begin{equation}
\bms{\hat E}(\bms{k},t) = \Big( \bms{D}_1(\bms{k})+\bms{D}_2
(\bms{k}) \Big) \hat{Q}(\bms{k},t) =
-4\pi \bms{\hat P}_{\rm L}(\bms{k},t) D(k) \ ,
\end{equation}
where
$\hat Q(\bms{k},t)=\sum_{i, a}^{N,M} q \scs{a} {\mbox{\large e}}^{-{\rm i}
\bvs{k\!\cdot\!r}_i^a(t)}$, $\bms{\hat P}_{\rm L}(\bms{k},t)=\frac{{\rm i}
\bvs{k}}{k^2} \hat Q(\bms{k},t)$ is the longitudinal component of the
microscopic operator $\bms{\hat P}$ of polarization density ($\bms{\nabla}
\bms{\cdot} \bms{\hat P}(\bms{r},t)=-\hat Q(\bms{r},t)$) and
$D(k)=D_1(k)+D_2(k)$.

Let us apply an external electric field $\bms{E} \scs{0}(\bms{k},\omega)$
to the system under consideration. The longitudinal, wavevector- and
frequency-dependent dielectric constant is defined via the material
relation $4\pi \bms{P}_{\rm L}(\bms{k},\omega)=\Big(\varepsilon \scs{\rm L}
(k,\omega) - 1 \Big) \bms{E}_{\rm L}(\bms{k},\omega)$, where $\bms{P}_{\rm
L}(\bms{k},\omega)=\left< \bms{\hat P}_{\rm L}(\bms{k},\omega) \right>$
and $\bms{E}_{\rm L}(\bms{k},\omega)=\left< \bms{\hat k} \bms {\hat k}
\bms{\cdot} \bms{E} \scs{0}(\bms{k},\omega)+\bms{\hat E}(\bms{k},\omega)
\right>$ are macroscopic values for longitudinal components of the
polarization and total field, $\left< \ \ \right>$ denotes statistical
averaging at the presence of the external field and the time Fourier
transform ${\scr F}(\bms{k},\omega)={\displaystyle \int_{-\infty}^{\infty}}
{\rm d}t \, {\mbox{\large e}}^{-{\rm i} \omega t} \ {\scr F}(\bms{k},t)$ has
been used for the functions $\bms{\hat P}_{\rm L}(\bms{k},t)$, $\bms{\hat E}
(\bms{k},t)$ and $\bms{\hat k}=\bms{k}/k$. Perturbation theory of the first
order with respect to $\bms{E} \scs{0}$ yields $\bms{P}_{\rm L}(\bms{k},
\omega)=- \frac{1}{V k_{\rm B} T} {\displaystyle \int_{0}^{\infty}} {\rm d}
t \, {\mbox{\large e}}^{-{\rm i} \omega t} \frac{{\rm d}}{{\rm d} t} \left<
\bms{\hat P}_{\rm L}(\bms{k},0) \bms{\cdot} \bms{\hat P}_{\rm L}(-\bms{k},t)
\right> \scs{0} \bms{\hat k} \bms{\hat k} \bms{\cdot} \bms{E} \scs{0}
(\bms{k},\omega)$, where $\left< \ \ \right> \scs{0}$ denotes equilibrium
averaging at the absence of the external field, and $k_{\rm B}$ and $T$ are
the Boltzmann's constant and the temperature of the system, respectively.
Then, eliminating $\bms{E} \scs{0}(\bms{k},\omega)$ from the presented above
expressions, we obtain the desired fluctuation formula
\begin{equation}
\frac{\varepsilon \scs{\rm L}(k,\omega) - 1}
{\varepsilon \scs{\rm L}(k,\omega)}=\frac{9y
{\scr L}_{{\rm i}\omega} \left(-\dot{G}_{\rm L}(k,t) \right)}
{1+9y (1-D(k)) {\scr L}_{{\rm i}\omega} \left(-\dot{G}_{\rm L}(k,t)
\right)} = 9y {\scr L}_{{\rm i}\omega} \left(-\dot{g} \scs
{\rm L}(k,t) \right) \, .
\end{equation}
Here $G_{\rm L}(k,t)= \left<\bms{\hat P}_{\rm L}(\bms{k},0) \bms{\cdot}
\bms{\hat P}_{\rm L}(-\bms{k},t) \right> \scs{0} \Big/ N \mu^2$ is the
longitudinal component of the wavevector-dependent dynamical Kirkwood
factor for the finite system, $\mu=|\bms{\mu}_i|$ denotes the permanent
magnitude of the molecule's dipole moment $\bms{\mu}_i=\sum_a^M q \scs{a}
\bms{r}_i^a$, $y=4\pi N \mu^2 \Big/ 9Vk_{\rm B}T$ and ${\scr L}_{{\rm i}
\omega} \left( ... \right) = {\displaystyle \int_{0}^{\infty}} ... \
{\mbox{\large e}}^{-{\rm i} \omega t} {\rm d} t$ is the Laplace transform.
The right-hand side of Eq.~(5) corresponds to the well-known fluctuation
formula for infinite systems, where $g \scs{\rm L}(k,t)=\lim_{N \to \infty}
G_{\rm L}(k,t)$ is the infinite-system Kirkwood factor.

The computer adapted formula (5) reduces to the formula for infinite systems
if the function $D(k)=1$. It can be shown easily that for nonzero wavevectors
the function $D(k) \to 1$ if $k_{\rm max} \to \infty$, additionally provided
$\eta \to \infty$ at ${\cal N}=0$. For $k=0$ the pattern is different because
of finiteness of $L$ and $D(0)=0$ as in the ISRF geometry [22]. However, in
the case of an actual summation, when $k_{\rm max}$ takes finite values, the
factor $D(k)$ can noticeably differ from unity. Therefore, the finite sample
behaves like a macroscopic system if the function $D(k)$ is very close to
unity and this condition can be verified now quantitatively. Moreover, this
explicit result may serve as an initial point for a more fruitful discussion
about the Ewald method itself. Let $\Delta={\rm max}_k|1-D(k)|$ be a maximal
deviation of $D(k)$ from unity in the whole interval of acceptable nonzero
wavevector values for a chosen pair of parameters $\eta$ and $k_{\rm max}$.
Then an optimal value for $\eta$ can be determined as that providing a
global minimum for the function $\Delta(\eta, k_{\rm max})$ at a given
$k_{\rm max}$.

According to Eq.~(5), the obtained in simulations Kirkwood factor $G_{\rm L}$
differs from its genuine value $g \scs{\rm L}$ with the relative precision
of $\chi=9y \Delta$. The function $\chi(\eta, n_{\rm max})$ is shown in
Fig.~1 as depending on $\eta$ at fixed values of $n_{\rm max}=k_{\rm max}
L/2\pi$. It has been calculated for the case of $R=L/2$ and $y=5.47$ that
corresponds to the thermodynamics point $\rho=m N/V=1$ g/cm$^3$, $T=293$ K
of the TIP4P model, where $m$ is the mass of water molecule. As we can see
from the figure, the precision of calculations of dielectric quantities in
computer experiment depends on Ewald parameters in a characteristic way. We
indicate the existence of the sharp minimum of $\chi$ at an arbitrary value
of $n_{\rm max}$. The curves of Fig.~1 can be useful to estimate the
possibility of a given simulation result to reproduce directly the
macroscopic dielectric behaviour of an IS system in an arbitrary
thermodynamics state, because then the function $\chi'=\chi\,y'/y$ is
simply rescaled, using the actual value of $y'$. From the last equality it
follows that the precision of calculations is better for systems with lower
particle densities $N/V$, molecular polarities $\mu$ and higher temperatures
$T$. It is obvious also that minimums of the functions $\Delta$ and $\chi$
with respect to Ewald parameters coincide between themselves.

The optimal pairs of values for $\eta$ and $n_{\rm max}$ at $R=L/2$ as well
as the corresponding values of the functions $\Delta$ and $\chi$ are selected
in Table~1. Choosing a criterion $\chi \mathop{<} \limits_{\mbox {\small
$\sim$}} 1\%$, we may ask that the formula for infinite systems might be
applied (at $k \ne 0$) and the influence of summation details can be
neglected in this case for which $G_{\rm L}(k,t)$ and $g \scs{\rm L}(k,t)$
are indistinguishable. It can be seen easily from the table that values of
$n_{\rm max} \ge 4$ satisfy this criterion if the parameter $\eta$ is chosen
optimally. The parameters $n_{\rm max}=5$, $\eta L=5.76$ and $R=L/2$ are
usually exploited in simulations [10]. For these values the relative
precision is $\chi=0.22\%$. However, choosing the optimal value $\eta L=
5.929$ at $n_{\rm max}=5$ instead of $\eta L=5.76$, we can reduce the
uncertainty up to $\chi=0.13\%$.

In the presented above consideration, the cut-off radius $R$ has been putted
to be half the basic cell length. Nevertheless, increasing $n_{\rm max}$,
the same precision of summation can be achieved also at smaller values of
$R$. Let $\eta$ and $n_{\rm max}$ correspond to the optimal parameters at
$R=L/2$. And now we choose a smaller value of the cut-off radius in the form
$R'=R/l$, where $l > 1$. Taking into account the fact that maximum deviations
of $D(k)$ from unity are always observed at $k=2\pi(n_{\rm max}+1)/L$, it
is easy to show that the same value of $\Delta(\eta, n_{\rm max})$ can be
obtained also at $\eta'=l \eta$ and $n_{\rm max}'=l (n_{\rm max}+1)-1$. For
example, putting $n_{\rm max}=5$ and $\eta L=5.929$ at $R=L/2$, we then
obtain for $R=L/4$ $(l=2)$ the following results: $\eta' L=11.858$ and
$n_{\rm max}'=11$. Choosing smaller values of $R$ can be more convenient
if the summation in $\bms{k}$-space takes less computation time in an actual
programme than the summation in real space. Indeed, let $t_1$ and $t_2$ are
the computation times in real and $\bms{k}$-space ($t_1>t_2$), respectively,
at given values of $R$ and $n_{\rm max}$. It is obvious that $t_1 \sim R^3$
and $t_2 \sim (n_{\rm max}+1)^3$. Then using new values $n_{\rm max}'=l
(n_{\rm max}+1)-1$ and $R'=R/l$ and minimizing the total computation time
$t'=t_1'+t_2'$ with respect to $l$, one obtains $l=\sqrt[6]{t_1/t_2}$.
Therefore, in such a way we can provide even a time optimization of
the programme without any loss of the precision.

\vspace{12pt}

\section{Numerical results. Comparing the Ewald and reaction field
methods}

\vspace{6pt}

\hspace{1em}  The study of dielectric properties by computer experiment is
still a major challenge, given that the calculations are very sensitive to
long-range interactions and because the polarization of polar fluids is a
collective effect, so that long trajectories are required in order to obtain
adequate statistical accuracy. For this reason, until now, the dynamical
polarization of ISMs has been investigated at zero or small wavevector
values only [18, 19, 23, 24, 27, 29, 31]. As far as we know, there are no
computer experiment data on the entire wavevector dependence of dynamical
dielectric quantities for such systems.

Our molecular dynamics simulations were carried out for the TIP4P model [32]
in the microcanonical ensemble at a density of $\rho=1$ g/cm$^3$ and at a
temperature of $T=293$ K. We have performed two runs corresponding the Ewald
and ISRF [22] geometries, respectively. In the both runs $N=256$ molecules
were considered in the cubic sample $V=L^3$ to which toroidal boundary
conditions were applied (${\cal N}=0$) and the interaction cut-off radius
was half the basic cell length, $R=L/2=9.856$\AA. The simulations were
started from a well equilibrated configuration for positions of sites,
obtained by Monte Carlo simulations. Initial velocities of molecules were
generated at random with the Maxwell distribution. The equations of motion
were integrated with a time step of $\Delta t=2$ fs on the basis of a matrix
method [33] using the Verlet algorithm in velocity form. The system was
allowed to achieve equilibrium for 50 000 time steps. The equilibrium state
was observed during 500 000 $\Delta t=1$ ns and each 10th time step was
chosen to compute equilibrium averages. Translational and angular velocities
of molecules were slightly rescaled after every 500 time steps in order to
conserve the total energy of the system, so that the relative total energy
fluctuations did not exceed 0.01\% over the whole runs.

The dynamical Kirkwood factor was evaluated in the time interval of
$1000 \Delta t=2$ ps and in a very large wavenumber region, namely, at
$k=[0,1,\ldots,300] k_{\rm min}$, where $k_{\rm min}=2 \pi/L=0.319{\rm
\AA}^{-1}$. Considering the system during such a rather long period of
time allows us to achieve statistical accuracy for the investigated
quantities of order 1\%. The optimal parameters $\eta L=5.929$ and
$n_{\rm max} = 5$ have been used in the Ewald summation of Coulomb
forces. The computational times on IBM PC AT486DX4 100 MHz to evaluate
dynamics of the system in our Fortran programmes were 2.2 s and 1.2 s
per step in the cases of Ewald and ISRF geometries, respectively.

Within the Ewald geometry the Coulomb part $q \scs{a} q \scs{b}
/|\bms{\rho}_{ij}^{ab}|$ of the intersite potential is replaced by
\begin{equation}
\varphi \scs{ij}^{ab} \!=\! q \scs{a} q \scs{b} \Bigg\{
\Theta \Big(R-|\bms{\overline \rho}_{ij}^{ab}|\Big)
\frac{{\rm erfc}(\eta |\bms{\overline \rho}_{ij}^{ab}|)}
{|\bms{\overline \rho}_{ij}^{ab}|}
+ \frac{4\pi}{V} \sum_{|\bvs{k}|>0}^{k_{\rm max}} 
\frac{  {\mbox{\large e}}^{-\frac{k^2}{4\eta^2}} }{k^2}
\cos(\bms{k} \bms{\cdot} \bms{\rho}_{ij}^{ab}) \Bigg\}
\!=\!\varphi_1(|\bms{\overline \rho}_{ij}^{ab}|)+
\varphi_2(\bms{\rho}_{ij}^{ab}) \, . \\ [4pt]
\end{equation}
Here, $\bms{\rho}_{ij}^{ab}=\bms{r}_i^a-\bms{r}_j^b$ designates the distance
between sites belonging the basic cell ($\bms{r}_i^a, \bms{r}_j^b \in V$),
$\bms{\overline \rho}_{ij}^{ab}=\bms{r}_i^a-\bms{\overline r}_j^b$, where,
according to the toroidal boundary conditions, $\bms{\overline r}_j^b=
\bms{r}_j^b+\bms{p}L$ $(\bms{p}=(p_x,p_y,p_z)$; $p_x,p_y,p_z=0,\pm1$)
is the position of the nearest image of $\bms{r}_j^b$ with respect to
$\bms{r}_i^a$, and $\Theta$ denotes the Heviside function, i.e., $\Theta
(\rho)=1$ if $\rho \ge 0$ and $\Theta(\rho)=0$ otherwise. The function
$\Theta$ indicates about the spherical site-site truncation in the real
coordinate space. The force acting on the $a$\,-th charged site of molecule
$i$ due to the interaction with the $b$\,-th charge of molecule $j$ is
$\bms{F}_{ij}^{ab} = - \partial \varphi \scs{ij}^{ab}/\partial \bms{r}_i^a$
or in a more explicit form
\begin{eqnarray}
\bms{F}_{ij}^{ab} = q \scs{a} q \scs{b} \Bigg\{
\frac{\bms{\overline \rho}_{ij}^{ab}}{|\bms{\overline \rho}_{ij}^{ab}|^3}
\Theta \Big(R-|\bms{\overline \rho}_{ij}^{ab}|\Big) \left[
{\rm erfc}(\eta |\bms{\overline \rho}_{ij}^{ab}|)
+ \frac{2\eta}{\sqrt{\pi}} |\bms{\overline \rho}_{ij}^{ab}|
{\mbox{\large e}}^{-\eta^2 |\bvs{\overline \rho}_{ij}^{ab}|^2} \right]
\nonumber \\ [-8pt] \\
+ \frac{4\pi}{V} \sum_{|\bvs{k}|>0}^{k_{\rm max}} \frac{\bms{k}}{k^2}
{\mbox{\large e}}^{-\frac{k^2}{4\eta^2}} \sin(\bms{k} \bms{\cdot}
\bms{\rho}_{ij}^{ab}) \Bigg\} \equiv q \scs{a} q \scs{b} \Big(
\bms{D}_1(|\bms{\overline \rho}_{ij}^{ab}|)+
\bms{D}_2(\bms{\rho}_{ij}^{ab}) \Big) \ . \nonumber
\end{eqnarray}
We note that the $\delta$-like part $q \scs{a} q \scs{b} \frac{\bms{\overline
\rho}_{ij}^{ab}}{R^2} \delta(R-|\bms{\overline \rho}_{ij}^{ab}|) {\rm erfc}
(\eta R)$ of the force is not included in (7) for the reason that the
complementary error function vanishes at $|\bms{\rho}_{ij}^{ab}|=R$ for
sufficiently large values of $\eta$. In particular, in our case ${\rm erfc}
(\eta R)={\rm erfc}(\eta L/2)=0.0000276 \ll 1$. Then the potential (6) can
be considered as a continuous and continuously differentiable one and the
drift of the total energy of the system, associated with the passage of
sites through the surface of the truncation sphere, can be neglected.

In an actual molecular dynamics programme the current potential energy of the
system, $U=\frac12 \sum_{i \ne j}^{N} \sum_{a,b}^M \varphi \scs{ij}^{ab}$,
and the total force acting on the $a$\,-th site of molecule $i$ due to
interactions with all the rest of sites belonging other molecules,
$\bms{F}_{i}^{a} = \mathop{\sum_{j=1}^N} \limits_{(j \ne i)} \sum_{b=1}^M
\bms{F}_{ij}^{ab}$, can be calculated as follows
\begin{equation}
U = \frac12 \left\{ \sum_{i \ne j}^N \sum_{a,b}^M
\varphi_1(|\bms{\overline \rho}_{ij}^{ab}|) + \frac{1}{\pi L}
\sum_{|\bvs{n}|>0}^{n_{\rm max}} \frac{{\mbox{\large e}}^{-\frac{\pi^2 n^2}
{\eta^2 L^2}}}{n^2} {\scr Y}(\bms{n}) {\scr Y}(\bms{-n}) -
\sum_{i=1}^N \sum_{a,b}^M \varphi_2(\bms{\rho}_{ii}^{ab})
\right\} \, , \\ [-6pt]
\end{equation}
\begin{equation}
\bms{F}_{i}^{a} \!= q \scs{a} \! \mathop{\sum_{j=1}^N} \limits_{(j \ne i)}
\sum_{b=1}^M q \scs{b} \bms{D}_1(|\bms{\overline \rho}_{ij}^{ab}|)
+ \frac{2 q \scs{a}}{L^2} \! \sum_{|\bvs{n}|>0}^{n_{\rm max}}
\frac{\bms{n}}{n^2} {\mbox{\large e}}^{-\frac{\pi^2 n^2}{\eta^2 L^2}}
{\rm i} {\mbox{\large e}}^{-2\pi {\rm i} \bvs{n} \bvs{\cdot} \bvs{r}_i^a/L}
{\scr Y}(\bms{-n}) - q \scs{a} \sum_{b=1}^M q \scs{b}
\bms{D}_2(\bms{\rho}_{ii}^{ab}) \ , \\ [2pt]
\end{equation}
where the self electrostatic energy, $u = \frac12 \sum_{i=1}^N \sum_{a
\ne b}^M \varphi \scs{ii}^{ab}$, and the self forces, $\bms{f}_i^a =
{\mathop{\sum_{b=1}^M} \limits_{(b \ne a)}} \bms{F}_{ii}^{ab}$, have been
excluded from (8) and (9), respectively, because the intramolecular forces
do not contribute into molecular translational accelerations and torques.
The auxiliary function
\begin{equation}
{\scr Y}(\bms{n})=
\sum_{i,a}^{N,M} q\scs{a}
{\mbox{\large e}}^{-2\pi {\rm i} \bvs{n} \bvs{\cdot} \bvs{r}_i^a/L}
={\rm Re} {\scr Y}(\bms{n}) + {\rm i}\,{\rm Im} {\scr Y}(\bms{n})
\end{equation}
is introduced in order to reduce the total number of numerical operations
in $\bms{k}$-space from of order $(N M)^2 n_{\rm max}$ to $N M n_{\rm max}$
that is very important for simulating large systems. This number can be
reduced approximately twice yet, using invariance of the subsume expressions
with respect to the inverse transformation $\bms{n} \to \bms{-n}$. Finally,
taking into account that the real part of ${\scr Y}(\bms{n})$ is an even
function of $\bms{n}$ and the imaginary part is an odd one, we obtain that
only the real part ${\rm Re}\,{\rm i} {\mbox{\large e}}^{-2\pi {\rm i}
\bvs{n} \bvs{\cdot} \bvs{r}_i^a/L} {\scr Y}(\bms{-n}) = \sin(2\pi \bms{n}
\bms{\cdot} \bms{r}_i^a/L) {\rm Re} {\scr Y}(\bms{n}) + \cos(2\pi \bms{n}
\bms{\cdot} \bms{r}_i^a/L) {\rm Im} {\scr Y}(\bms{n})$ give nonzero
contributins into (9) and ${\scr Y}(\bms{n}) {\scr Y}(\bms{-n})=({\rm Re}
{\scr Y}(\bms{n}))^2 + ({\rm Im} {\scr Y}(\bms{n}))^2$.

In the RF geometry, real particles of the infinite system, which are
located outside the sphere of finite radius $R$ around a reference
particle belonging the basic cell, are replaced by an infinite, as a
rule, conducting continuum. There are two versions of the RF geometry.
In the PDRF approach, molecules are considered as point dipole particles
and the intermolecular potential is of the form [18, 19]:
\begin{equation}
\varphi \scs{ij}^{\rm PD} = \Theta (R-|\bms{r}_i-\bms{\overline r}_j|)
\left( \sum_{a,b}^M
\frac{q \scs{a} q \scs{b}}{|\bms{\overline \rho}_{ij}^{ab}|} -
\frac{\bms{\mu}_i \bms{\cdot} \bms{\mu}_j}{R^3} \right) \, ,
\end{equation}
where $\bms{r}_i$ is the centre of mass of the $i$-th molecule and the
molecular cut-off is performed. In the exact ISRF method [22] the spatial
distribution of charges within the molecule is taken into account explicitly
at constructing the reaction field. As a result, the potential (11)
transforms into
\begin{equation}
\varphi \scs{ij}^{\rm RF} = \sum_{a,b}^M q \scs{a} q \scs{b}
\Theta \Big(R-|\bms{\overline \rho}_{ij}^{ab}|\Big) \left\{ \frac{1}
{|\bms{\overline \rho}_{ij}^{ab}|} + \frac12 \frac{|\bms{\overline
\rho}_{ij}^{ab}|^2}{R^3} - \frac{3}{2R} \right\} \, ,
\end{equation}
where the first term in the right-hand side of (12) describes the usual
Coulomb field, whereas the rest of terms corresponds to the reaction field
in the IS description.

It can be shown easily that the potential (12) is reduced to (11) in one
case only, namely, when $d/R \to 0$, where $d=2 \max_a|\bms{r}_i^a-
\bms{r}_i|$ denotes the diameter of the molecule. In this case, the
positions for sites and centres of mass are undistinguished within
the same molecule. For finite samples of IS molecules we have $d/R \ne
0$ and, therefore, the PDRF potential (11) may affect on a true macroscopic
behaviour of the system considerably. Moreover, the ISRF method has yet a
minor advantage over the PDRF scheme that the potential of interaction (12)
is continuous and continuously differentiable. It is worth to mention also
that in the RF geometry the dielectric permittivity is computed using the
fluctuation formula (5) with the formal substitution $D(k) \to D_{\rm RF}(k)
= 1-3 j \scs{1}(kR)/(kR)$ [22].

The wavevector-dependent static Kirkwood factor, $G_{\rm L}(k) \equiv G_{\rm
L}(k,0)$, and samples of the normalized dynamical Kirkwood factor, $\Phi_{\rm
L}(k,t)=G_{\rm L}(k,t)/G_{\rm L}(k)$, calculated in the simulations within
the Ewald and ISRF geometries, are shown in Figs.~2, 3 by the circles
and dashed curve, respectively. Since in the ISRF geometry the function
$D_{\rm RF}(k)$ differs from unity considerably, to evaluate the infinite
system Kirkwood factor $g \scs{\rm L}(k,t)$ the performance of the
self-consistent transformation (5) is necessary. This result is plotted by
the solid curve. At the same time, within the Ewald geometry the function
$D(k)$ is very close to unity at the given optimal parameters of summation
(see Table~1), so that the infinite system Kirkwood factor is equivalent to
that, obtained directly in the simulations, i.e, $g \scs{\rm L}(k,t) =
G \scs{\rm L}(k,t)$ (excepting the case $k=0$). As we can see from the
figures, the agreement between the two sets of data for the infinite-system
functions, corresponding to the Ewald and ISRF geometries, is quite good.
The slight difference (within a few per cent) at large times can be
explained by an approximate character of the integration appearing for
the ISRF geometry at performing the inverse Laplace transform of (5).

For the purpose of comparison, the infinite-system Kirkwood factor
$g \scs{\rm L}(k)$ corresponding to the PDRF geometry is also included in
the Fig.~2 (the dotted curve). Deviations of values for $g \scs{\rm L}(k)$
obtained using the PDRF potential from those evaluated in the Ewald and
ISRF geometries are of order $20\%$. They are well exhibited at intermediate
values of wavevectors. Such a situation can be explained by the fact that
the PDRF geometry does not take into account the spatial distribution of
charges within the molecule and, thus, the precision of calculations for
wavevector-dependent dielectric quantities at $k \sim 2 \pi/d \sim
3.4{\rm \AA}^{-1}$ can not exceed $d/R \sim 20\%$, where $d=1.837{\rm
\AA}$ for the TIP4P water molecule. And only for great wavevector values
$(k>6{\rm \AA}^{-1})$, where the influence of boundary conditions is
negligible ($D_{\rm RF}(k) \to 1$), all the three geometries become
completely equivalent.

\vspace{12pt}

\section{Conclusion}

\hspace{1em}  Explicitly considering details of the Ewald summation to
treat Coulomb interactions, the fluctuation formula for the computation
of the dielectric permittivity in IS models of polar fluids has been
rigorously derived. Using this formula, it has been corroborated by actual
molecular dynamics calculations that the Ewald and ISRF methods can be
applied with equal successes to investigate the dielectric constant of ISMs
in computer experiment. The Ewald geometry, however, at a specific choice
for parameters of the summation, may reproduce the macroscopic behaviour
for dielectric quantities directly in simulations without any additional
transformations.

Since the calculation of the wavevector- and frequency-dependent dielectric
permittivity in simulations for ISMs is practical now in principle, we
believe that this fact will stimulate further research of such systems
in theory, computer and pure experiment.

\vspace{18pt}

{\bf Acknowledgements.} The author would like to acknowledge financial
support of the President of Ukraine.

\newpage

\vspace*{1cm}

{\bf Table 1.} Optimal parameters of the Ewald summation for ISMs at $R=L/2$

\vspace{0.25cm}

\begin{displaymath}
\begin{array}{||c|c|c|c||}
\hline
\hline & & & \\ [-12pt]
\ \ \ \ \ \eta L \ \ \ \ \ & \ \ \ n_{\rm max} \ \ \ &
\ \ \ \ \ \Delta (\%) \ \ \ \ \ & \ \ \ \chi (\%) \ \ \ \\ [12pt]
\hline
\hline & & & \\ [-12pt]
     3.301   &   1   &   1.632{\rm E}\!+\!00   &    80.29         \\
     3.874   &   2   &   5.523{\rm E}\!-\!01   &    27.17         \\
     4.791   &   3   &   6.684{\rm E}\!-\!02   &     3.288        \\
     5.209   &   4   &   2.212{\rm E}\!-\!02   &     1.088        \\
     5.929   &   5   &   2.690{\rm E}\!-\!03   &     0.1324       \\
     6.276   &   6   &   8.978{\rm E}\!-\!04   &     0.0442       \\
     6.887   &   7   &   1.094{\rm E}\!-\!04   &     0.0054       \\
     7.251   &   8   &   3.602{\rm E}\!-\!05   &     0.0018       \\ [-12pt]
& & & \\
\hline
\hline
\end{array}
\end{displaymath}

\vspace{2.5cm}

\begin{center}
{\large Figure captions}
\end{center}

\vspace{6pt}

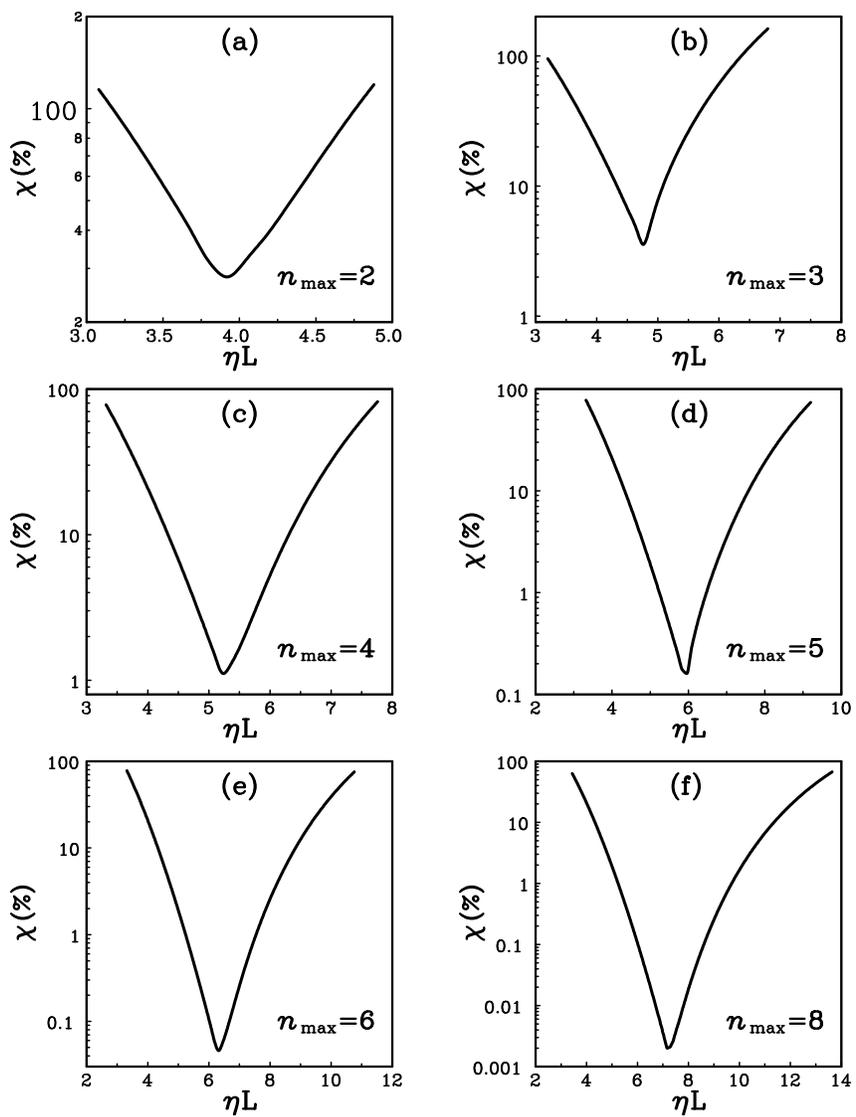
{\bf Fig.~1.}~The precision of reproducing bulk dielectric quantities
in computer experiment for ISMs as depending on parameters of the Ewald
geometry.

\vspace{6pt}

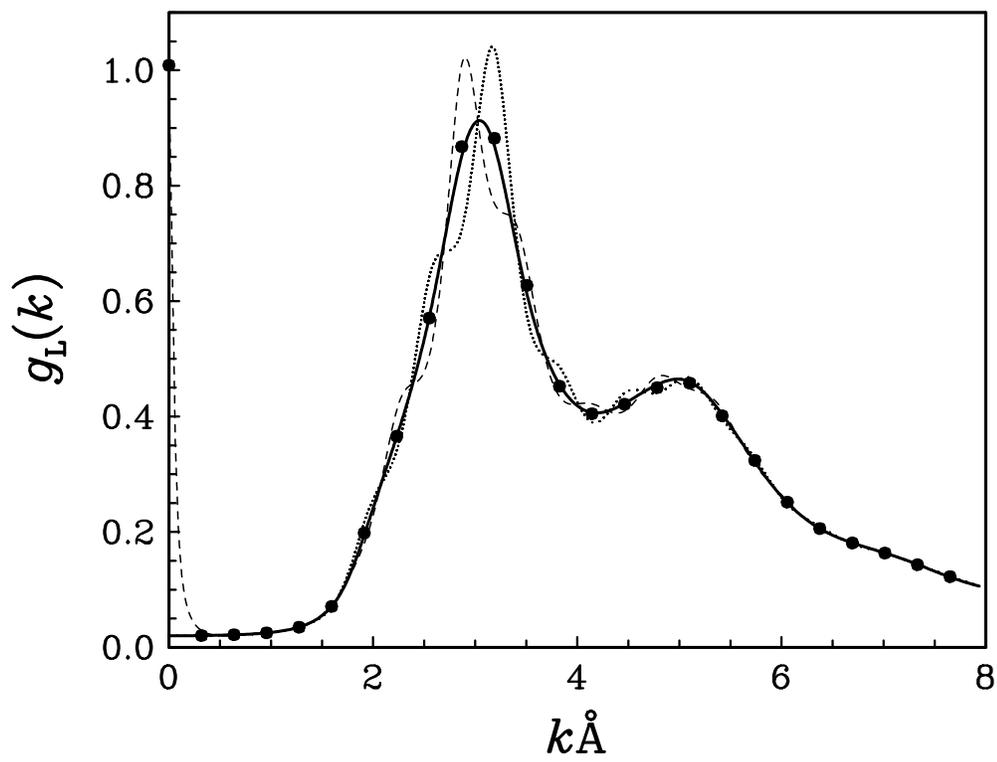
{\bf Fig.~2.}~The static wavevector-dependent Kirkwood factor of the TIP4P
water. The obtained result in the Ewald geometry is presented by the full
circles. The dashed and solid curves correspond to the finite and infinite
systems in the ISRF geometry. The PDRF infinite-system Kirkwood factor
is shown as the dotted curve.

\vspace{6pt}

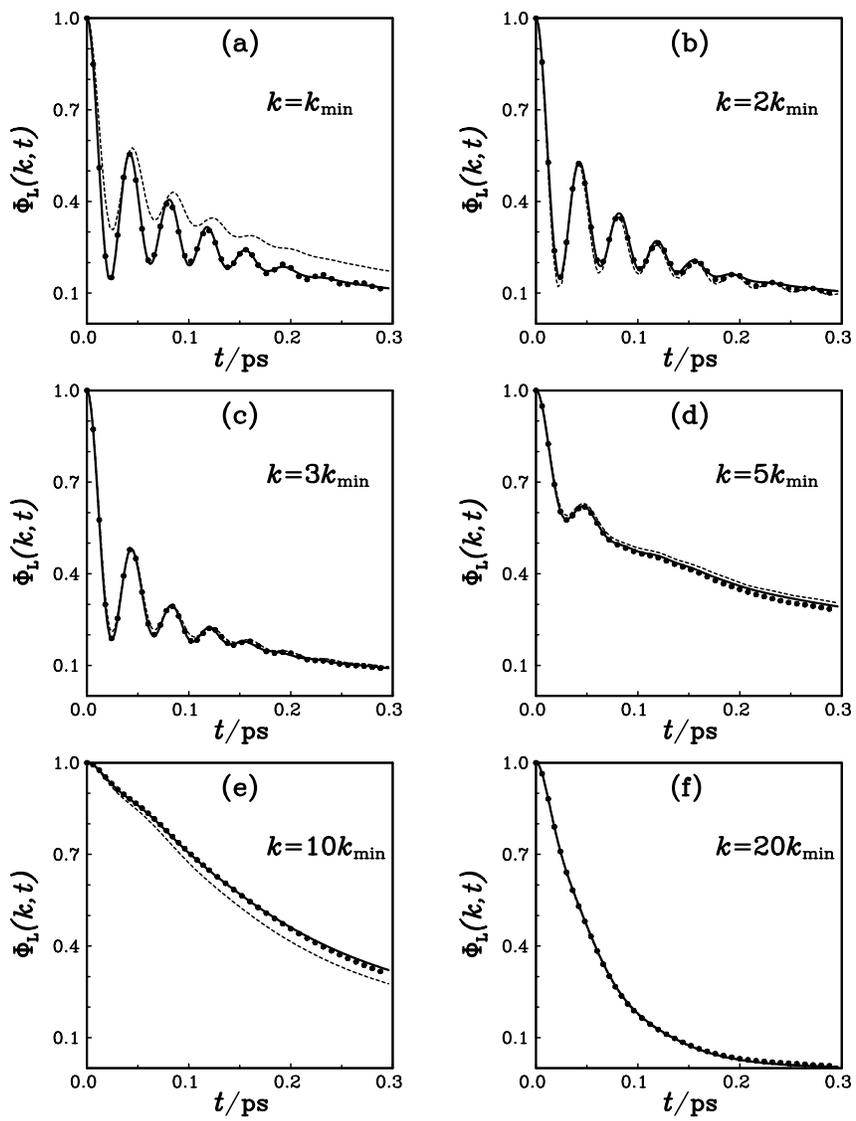
{\bf Fig.~3.}~The normalized dynamical Kirkwood factor of the TIP4P water.
Notations as for fig.~2.

\end{document}